\documentclass[3p,times,procedia]{elsarticle}
\usepackage{nupha_ecrc, hyperref}

\volume{00}

\firstpage{1}

\journalname{Nuclear Physics A}

\runauth{}

\jid{nupha}

\jnltitlelogo{Nuclear Physics A}

\usepackage{amssymb}
\usepackage[figuresright]{rotating}

\newcommand{\trento}{T$\mathrel{\protect\raisebox{-2.1pt}{R}}$ENTo}

\begin{document}

\begin{frontmatter}



%% Instructions from Editor: Please use the following \dochead only in the preprint version (e-print arXiv etc.); 
%% use empty \dochead{} when submitting to Nuclear Physics A!
\dochead{XXVIIth International Conference on Ultrarelativistic Nucleus-Nucleus Collisions\\ (Quark Matter 2018)}
%% e.g. \dochead{17th International Conference on Dynamical Processes in Excited States of Solids}

\title{Confronting hydrodynamic predictions with Xe-Xe data}
%\title{Hydrodynamic behavior confronts Xe-Xe data}


\author[Saclay]{Giuliano Giacalone}
\author[Rutgers]{Jacquelyn Noronha-Hostler}
\author[USP]{Matthew Luzum}
\author[Saclay]{Jean-Yves Ollitrault}

\address[Saclay]{Institut de physique th\'eorique, Universit\'e Paris Saclay, CNRS, CEA, F-91191 Gif-sur-Yvette, France}
\address[Rutgers]{Department of Physics and Astronomy, Rutgers University,
	Piscataway, NJ 08854, USA}
\address[USP]{Instituto de F\'{i}sica, Universidade de S\~{a}o Paulo, C.P.
	66318, 05315-970 S\~{a}o Paulo, SP, Brazil}

\begin{abstract}
Comparing collision systems of different size, at near the same collision energy, offers us the opportunity to probe the scaling behavior and therefore the nature of the system itself.  Recently, we made predictions for Xe-Xe collisions at 5.44 TeV using viscous hydrodynamic simulations, noting that the scaling from the larger Pb-Pb system is rather generic, and arguing that robust predictions can be made that do not depend on details of the model.  Here we confront our predictions with measurements that were subsequently made in a short Xe-Xe run at the LHC by the ALICE, ATLAS, and CMS collaborations.  We find that the predictions are largely confirmed.
%, with small discrepancies that could point the way to a better understanding of the medium created in such collisions.   
Of particular interest is a strong indication of a non-spherical shape for the $^{129}$Xe nucleus.
\end{abstract}


\end{frontmatter}

% \linenumbers

\section{Introduction}
\label{}
In the standard picture of a heavy-ion collision, the system evolves as a relativistic fluid for a significant part of its lifetime.  Experimental data show many generic (qualitative and quantitative) features supporting this picture, and hydrodynamic simulations have successfully described and predicted a large set of observables.  Nevertheless, it is important to take advantage of every opportunity to test this standard picture as strictly as possible.   This can be accomplished by changing aspects of the  system, such as selecting events for collision centrality and changing the collision energy.   

Of particular interest is the scaling of results with system size. Recent measurements from small collision systems (such as p-p, p-A, d-A, and $^3$He-A) show many features that are similar to the largest collision systems.  This has lead to a vigorous debate about the nature of small collision systems as well as large, and the limits of validity of the equations of hydrodynamics.  

A natural way to address these questions is to perform collisions of intermediate size and verify whether the scaling is as predicted from the generic properties of hydrodynamic evolution.  To this end a short run was performed at the Large Hadron Collider using $^{129}$Xe beams at $\sqrt{s_{NN}}$ = 5.44 TeV, bridging the gap between the larger system of $^{208}$Pb-$^{208}$Pb and smaller systems p-p and p-$^{208}$Pb, which have been studied at almost the same energy (5.02 TeV).

In Ref.~\cite{Giacalone:2017dud} we made predictions for the upcoming Xe-Xe results, focusing on observables that probe generic scaling properties of the standard hydrodynamic picture (rather than those which depend on unknown parameters and model details), to test the nature of the collision system and whether system size scaling deviates from hydrodynamic expectations.  
%In particular, we focus on the scaling of results from the Pb-Pb system to the smaller Xe-Xe system.

Additional predictions and comparisons were made using another hydrodynamic model in Refs.~\cite{Eskola:2017bup, Niemi:2018usq}.
\section{Hydrodynamic scaling predictions and comparison to measurement}
Simulations were carried out for both Pb-Pb and Xe-Xe collisions using the 2+1 dimensional code v-USPhydro \cite{Noronha-Hostler:2013gga,Noronha-Hostler:2014dqa}, assuming longitudinal boost invariance.  Initial conditions are given by the \trento\ \cite{Moreland:2014oya} model with parameters $p=0$, $k=1.6$, $\sigma = 0.51$ fm, which provide the initial entropy density at an initial time of $\tau$ = 0.6 fm.  The initial transverse flow and viscous tensor are assumed to vanish.  Hydrodynamic evolution is carried out with shear viscosity $\eta/s=0.047$ and lattice QCD equation of state
PDG16+/2+1[WB]~\cite{Borsanyi:2013bia}
until freeze out at $T=$150 MeV, followed by the decay of unstable resonances.  
Nuclear shape parameters for the $^{129}$Xe nucleus taken from Ref.~\cite{Moller:2015fba}.

A baseline expectation is scale-invariance, due to the scale invariance of the equations of ideal hydrodynamics.   In this baseline picture, intensive quantities such as the mean transverse momentum and the various integrated flow observables $v_n$ should be equal in the larger (Pb) an smaller (Xe) systems.  Various effects break this scaling, but they can be understood as corrections to the scale-invariant baseline.
The mean transverse momentum, for example, is expected to change very little (less than 2\%)\footnote{A measurement of $\langle p_T\rangle$ was presented by the ALICE collaboration in Ref.~\cite{Acharya:2018eaq}, but the result is under revision and we were asked not to compare to the current result.} \cite{Giacalone:2017dud}.
Flow measurements are instead expected to change more, due to the following effects.  

\begin{figure}[ht!]
	\begin{center}
		\includegraphics[width=\linewidth]{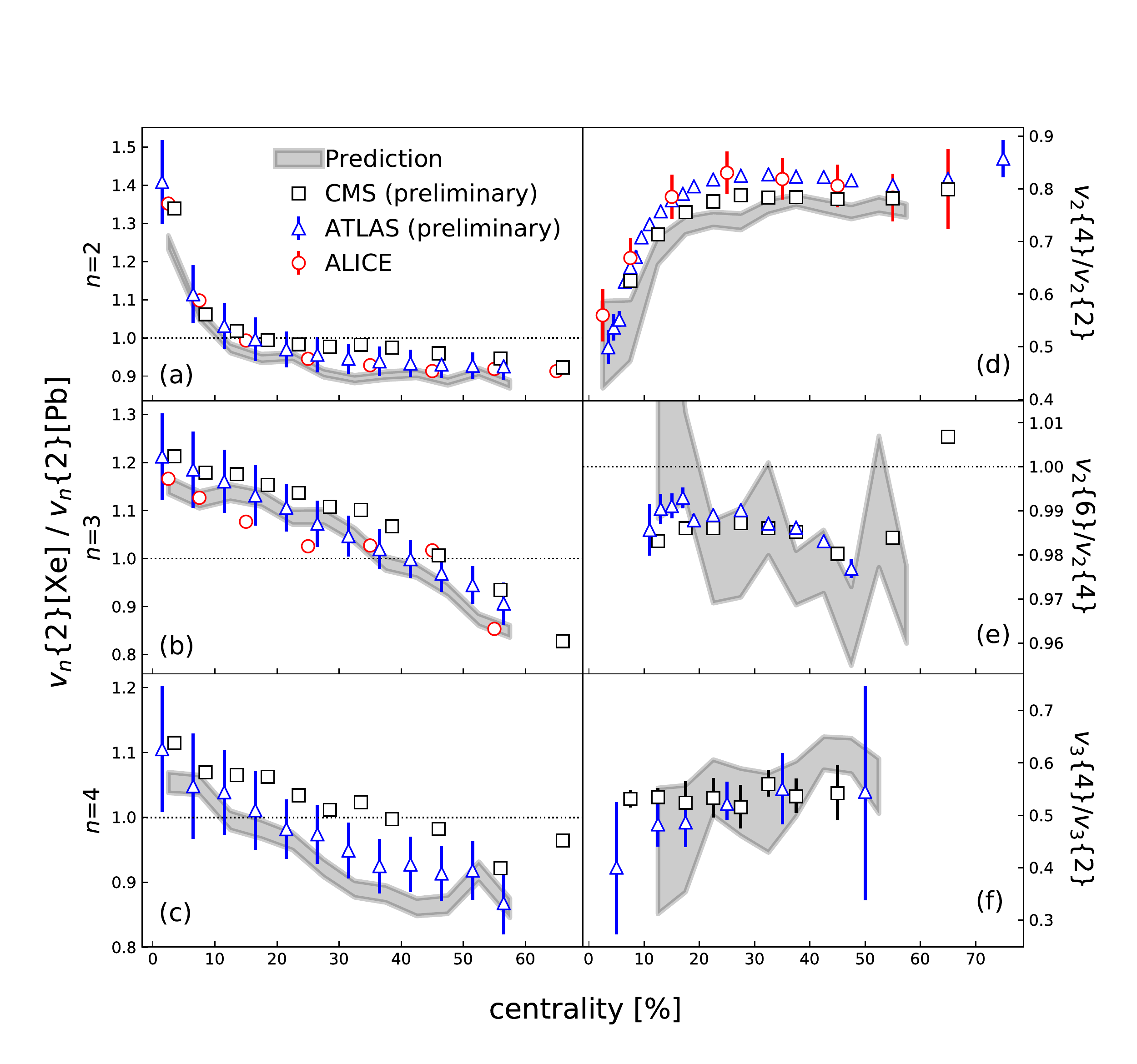}
	\end{center}
	%	\vspace*{-7mm}
	\caption{\label{fig}Hydrodynamic predictions (gray bands) compared to measurements from ALICE (circles) 
		%		\cite{Acharya:2018ihu, Acharya:2018eaq} 
		\cite{Acharya:2018ihu}
		and preliminary measurements from ATLAS (triangles) \cite{ATLAS:2018iom}, and CMS (squares) \cite{CMS:2018jmx}.  Shown on the left is the ratio of charged hadron $v_n\{2\}$ as measured in Xe-Xe collisions divided by that in Pb-Pb collisions for $v_2$ (a), $v_3$ (b), and $v_4$ (c).  For readability CMS points are shifted to the right 1\% and ATLAS to the left 1\%.   On the right are the ratio of charged hadron cumulants $v_2\{4\}/v_2\{2\}$ (d), $v_2\{6\}/v_2\{4\}$ (e), and $v_3\{4\}/v_3\{2\}$ (f) in Xe-Xe collisions
		%		, as well as the ratio of $\langle p_t\rangle$ of charged hadrons in Xe-Xe  to Pb-Pb (f).
	}
\end{figure}

Fluctuations in the initial state, instead of being scale invariant, are expected to scale as $A^{-1/2}$.  
This results in an expected \textit{increase} in the smaller system of $(129/208)^{-1/2}\sim 1.27$, 
 which is verified in calculations of triangularity $\epsilon_3$\cite{Giacalone:2017dud}.  However, the hydro response to the initial spatial anisotropy $v_3/\epsilon_3$ is also modified due to viscous effects, which are greater in the smaller system\footnote{Note that here, ``viscous effects'' refers not only to the value of viscosity in the Quark-Gluon-Plasma phase of hydrodynamic evolution, but also effects such as freeze out that are present even when viscosity is negligible near the transition temperature.  All such effects are expected to be larger for systems of smaller size}.  Overall, these effects combine for a predicted increase in $v_3$ in Xe-Xe collisions of approximately 15\% in central collisions, and a \textit{decrease} of $\sim$10\% in collisions at 50--60\% centrality. 
Measurements of $v_3\{2\}$ were made subsequently by the three major LHC experiments and are presented along with our prediction in Fig.~\ref{fig}b.  There is some disagreement between experimental measurements, but overall the prediction is quantitatively validated, indicating that the expected hydrodynamic scaling is present and also that the viscous effects in our calculation are of a realistic size.    A similar prediction was made for $v_4$, which is compared to preliminary results from the CMS and ATLAS collaborations in Fig.~\ref{fig}c.

Elliptic flow is generated not only from fluctuations in the initial stages of a collision, but also by the average geometry of the nuclear overlap region.   In non-central collisions, the latter is the dominant effect, and so a significant increase in eccentricity $\epsilon_2$ is not expected.  Instead, increased viscous effects (plus a slightly less sharp nuclear skin) predict a small decrease of $v_2$, in agreement with subsequent measurements, as seen in Fig.~\ref{fig}a.
In contrast fluctuations are important for central collisions.  Besides increased fluctuations due to the smaller size, the Xe system may exhibit additional elliptic flow fluctuations because of a non-spherical nuclear shape.  Such a deformation has not yet been observed in this isotope, but interpolation from nearby isotopes and theoretical calculations suggest that it should be present.  An observation of nuclear structure properties such as this in a heavy-ion collision would be very interesting.  
Indeed, measurements of $v_2\{2\}$ indicate a strong increase for central collisions (see Fig.~\ref{fig}a), as predicted for a deformed nucleus (but not for a spherical nucleus) \cite{Giacalone:2017dud}.  We know of no other explanation for this distinct dependence on centrality, and given the overall agreement with other harmonics and other centralities, we take it as strong evidence of a deformed $^{129}$Xe nucleus.

Cumulants of more than 2 particles are also of interest.  These observables provide additional information about fluctuations in the system.  For example, the ratio $v_2\{4\}/v_2\{2\}$ is directly related to the variance of the event-by-event distribution of $v_2^2$ --- when there are no event-by-event fluctuations, the ratio is 1, but is reduced as the magnitude of fluctuations is increased.  As such, we expect this ratio to be smaller in the Xe system compared to the Pb system.  This prediction was also verified by recent measurements, as seen in Fig.~\ref{fig}d.  (Note that our model has a ratio that is slightly too small in both systems, but the predicted change with system size is expected to be more robust, and is indeed correct). Similarly, preliminary data for the ratio $v_2\{6\}/v_2\{4\}$, shown in Fig.~\ref{fig}e, are in excellent agreement with our predictions, and they satisfy the expected scaling between Pb-Pb and Xe-Xe, in the sense that the deviation of this ratio from unity is larger in Xe-Xe systems. Since the degeneracy of higher-order cumulants is driven by nearly Gaussian $v_2$ fluctuations in presence of an almond shape of the collision zone \cite{Voloshin:2007pc}, one naturally expects Xe-Xe collisions, where fluctuations are larger and geometry is less sharp, to present a coarser splitting of higher-order cumulants. Finally, our prediction for the ratio $v_3\{4\}/v_3\{2\}$, shown in Fig.~\ref{fig}f, is in good agreement with preliminary data, though with larger error bars. No significant difference between Pb-Pb and Xe-Xe is observed for this observable.
%Subsequent measurements verified this prediction, as seen in Fig.~\ref{a}.
%
%\section{Results}
\section{Conclusions}
Hydrodynamic calculations were made for Pb-Pb and Xe-Xe collision systems, in order to make predictions for the smaller Xe system and systematically test the hydrodynamic paradigm via the scaling of observables with system size.  Here we compare to subsequent measurements, showing good agreement and a validation of the standard hydrodynamic picture across both systems.

A particular interesting result is the strong increase of $v_2\{2\}$ in central collisions in the Xe system compared to the Pb system, which is a distinct indication of non-spherical shape for the $^{129}$Xe nucleus.  Though a deformed shape was expected, LHC data provide the first experimental evidence, and this is arguably one of the most interesting and important results from the Xe-Xe run.
\section{Acknowledgments}
ML acknowledges support from FAPESP projects 2016/24029-6  and 2017/05685-2, and project INCT-FNA Proc.~No.~464898/2014-5.
J.N.H.~acknowledges the Office of Advanced Research Computing (OARC) at Rutgers, The State University of New Jersey for providing access to the Amarel cluster and associated research computing resources that have contributed to the results reported here. J.N.H.~also acknowledges the use of the Maxwell Cluster and the advanced support from the Center of Advanced Computing and Data Systems at the University of Houston as well as the support of the Alfred P. Sloan Foundation. 
This work was funded in part under the USP-COFECUB project Uc Ph 160-16
(2015/13) and under the FAPESP-CNRS project 2015/50438-8.



%% References


%\bibliographystyle{elsarticle-num}
%\bibliography{LuzumQM2018Proceedings}

\section*{References}
 \begin{thebibliography}{00}


%\cite{Giacalone:2017dud}
\bibitem{Giacalone:2017dud} 
G.~Giacalone, J.~Noronha-Hostler, M.~Luzum and J.~Y.~Ollitrault,
%``Hydrodynamic predictions for 5.44 TeV Xe+Xe collisions,''
Phys.\ Rev.\ C {\bf 97}, no. 3, 034904 (2018)
doi:10.1103/PhysRevC.97.034904
[arXiv:1711.08499 [nucl-th]].
%%CITATION = doi:10.1103/PhysRevC.97.034904;%%
%11 citations counted in INSPIRE as of 12 Jul 2018


%\cite{Eskola:2017bup}
\bibitem{Eskola:2017bup} 
K.~J.~Eskola, H.~Niemi, R.~Paatelainen and K.~Tuominen,
%``Predictions for multiplicities and flow harmonics in 5.44 TeV Xe+Xe collisions at the CERN Large Hadron Collider,''
Phys.\ Rev.\ C {\bf 97}, no. 3, 034911 (2018)
doi:10.1103/PhysRevC.97.034911
[arXiv:1711.09803 [hep-ph]].
%%CITATION = doi:10.1103/PhysRevC.97.034911;%%
%6 citations counted in INSPIRE as of 12 Jul 2018


%\cite{Niemi:2018usq}
\bibitem{Niemi:2018usq} 
H.~Niemi, K.~J.~Eskola, R.~Paatelainen and K.~Tuominen,
%``Latest predictions from the EbyE NLO EKRT model,''
arXiv:1807.02378 [nucl-th].
%%CITATION = ARXIV:1807.02378;%%

\bibitem{Noronha-Hostler:2013gga} 
  J.~Noronha-Hostler, G.~S.~Denicol, J.~Noronha, R.~P.~G.~Andrade and F.~Grassi,
  %``Bulk Viscosity Effects in Event-by-Event Relativistic Hydrodynamics,''
  Phys.\ Rev.\ C {\bf 88}, no. 4, 044916 (2013).
  %%CITATION = doi:10.1103/PhysRevC.88.044916;%%
  %112 citations counted in INSPIRE as of 13 Jul 2018
\bibitem{Noronha-Hostler:2014dqa} 
  J.~Noronha-Hostler, J.~Noronha and F.~Grassi,
  %``Bulk viscosity-driven suppression of shear viscosity effects on the flow harmonics at energies available at the BNL Relativistic Heavy Ion Collider,''
  Phys.\ Rev.\ C {\bf 90}, no. 3, 034907 (2014).
  %%CITATION = doi:10.1103/PhysRevC.90.034907;%%
  %73 citations counted in INSPIRE as of 13 Jul 2018


%\cite{Moreland:2014oya}
\bibitem{Moreland:2014oya} 
J.~S.~Moreland, J.~E.~Bernhard and S.~A.~Bass,
%``Alternative ansatz to wounded nucleon and binary collision scaling in high-energy nuclear collisions,''
Phys.\ Rev.\ C {\bf 92}, no. 1, 011901 (2015)
doi:10.1103/PhysRevC.92.011901
[arXiv:1412.4708 [nucl-th]].
%%CITATION = doi:10.1103/PhysRevC.92.011901;%%
%65 citations counted in INSPIRE as of 12 Jul 2018


%\cite{Borsanyi:2013bia}
\bibitem{Borsanyi:2013bia} 
S.~Borsanyi, Z.~Fodor, C.~Hoelbling, S.~D.~Katz, S.~Krieg and K.~K.~Szabo,
%``Full result for the QCD equation of state with 2+1 flavors,''
Phys.\ Lett.\ B {\bf 730}, 99 (2014)
doi:10.1016/j.physletb.2014.01.007
[arXiv:1309.5258 [hep-lat]].
%%CITATION = doi:10.1016/j.physletb.2014.01.007;%%
%403 citations counted in INSPIRE as of 12 Jul 2018


%\cite{Moller:2015fba}
\bibitem{Moller:2015fba} 
P.~Möller, A.~J.~Sierk, T.~Ichikawa and H.~Sagawa,
%``Nuclear ground-state masses and deformations: FRDM(2012),''
Atom.\ Data Nucl.\ Data Tabl.\  {\bf 109-110}, 1 (2016)
doi:10.1016/j.adt.2015.10.002
[arXiv:1508.06294 [nucl-th]].
%%CITATION = doi:10.1016/j.adt.2015.10.002;%%
%105 citations counted in INSPIRE as of 12 Jul 2018


%\cite{Acharya:2018eaq}
\bibitem{Acharya:2018eaq} 
S.~Acharya {\it et al.} [ALICE Collaboration],
%``Transverse momentum spectra and nuclear modification factors of charged particles in Xe-Xe collisions at $\sqrt{s_{\rm NN}}$ = 5.44 TeV,''
arXiv:1805.04399 [nucl-ex].
%%CITATION = ARXIV:1805.04399;%%
%1 citations counted in INSPIRE as of 12 Jul 2018


%\cite{Acharya:2018ihu}
\bibitem{Acharya:2018ihu} 
S.~Acharya {\it et al.} [ALICE Collaboration],
%``Anisotropic flow in Xe-Xe collisions at $\mathbf{\sqrt{s_{\rm{NN}}} = 5.44}$ TeV,''
doi:10.1016/j.physletb.2018.06.059
arXiv:1805.01832 [nucl-ex].
%%CITATION = doi:10.1016/j.physletb.2018.06.059;%%
%2 citations counted in INSPIRE as of 12 Jul 2018
J.~Margutti, these proceedings; K.~ Gajdosova, these proceedings


%\cite{ATLAS:2018iom}
\bibitem{ATLAS:2018iom} 
The ATLAS collaboration [ATLAS Collaboration],
%``Measurement of the azimuthal anisotropy of charged particle production in Xe+Xe collisions at $\sqrt{s_{\mathrm{NN}}}$=5.44~TeV with the ATLAS detector,''
ATLAS-CONF-2018-011.
%%CITATION = ATLAS-CONF-2018-011;%%
%1 citations counted in INSPIRE as of 12 Jul 2018
T.~Bold, these proceedings; I.~Grabowska-Bold, these proceedings


%\cite{CMS:2018jmx}
\bibitem{CMS:2018jmx} 
CMS Collaboration [CMS Collaboration],
%``Charged particle angular correlations in XeXe collision at $\sqrt{s_{NN}} = 5.44$ TeV,''
CMS-PAS-HIN-18-001.
%%CITATION = CMS-PAS-HIN-18-001;%%
M.~Stojanovic, these proceedings; M.~Verweij, these proceedings

%\cite{Voloshin:2007pc}
\bibitem{Voloshin:2007pc} 
  S.~A.~Voloshin, A.~M.~Poskanzer, A.~Tang and G.~Wang,
  %``Elliptic flow in the Gaussian model of eccentricity fluctuations,''
  Phys.\ Lett.\ B {\bf 659}, 537 (2008)
  doi:10.1016/j.physletb.2007.11.043
  [arXiv:0708.0800 [nucl-th]].
  %%CITATION = doi:10.1016/j.physletb.2007.11.043;%%
  %137 citations counted in INSPIRE as of 13 Jul 2018


 \end{thebibliography}



\end{document}